\documentclass[conference]{IEEEtran}
\IEEEoverridecommandlockouts
\usepackage{cite}
\usepackage{amsmath,amssymb,amsfonts}
\usepackage{algorithmic}
\usepackage{graphicx}
\usepackage{textcomp}
\usepackage{xcolor}
\def\BibTeX{{\rm B\kern-.05em{\sc i\kern-.025em b}\kern-.08em
    T\kern-.1667em\lower.7ex\hbox{E}\kern-.125emX}}
\begin{document}

\title{From Model-Based and Adaptive Control to Evolving Fuzzy Control}

\author{\IEEEauthorblockN{Daniel Leite}
\IEEEauthorblockA{\textit{University of Paderborn} \\
Paderborn, Germany \\
daniel.leite@uni-paderborn.de}
\and
\IEEEauthorblockN{Igor \v Skrjanc}
\IEEEauthorblockA{\textit{University of Ljubljana} \\
Ljubljana, Slovenia\\
igor.skrjanc@fe.uni-lj.si}
\and
\IEEEauthorblockN{Fernando Gomide}
\IEEEauthorblockA{\textit{University of Campinas} \\
Campinas, Brazil\\
gomide@unicamp.br}
}

\maketitle

\begin{abstract}

Evolving fuzzy systems build and adapt fuzzy models—such as predictors and controllers—by incrementally updating their rule-base structure from data streams. On the occasion of the 60-year anniversary of fuzzy set theory, commemorated during the Fuzz-IEEE 2025 event, this brief paper revisits the historical development and core contributions of classical fuzzy and adaptive modeling and control frameworks. It then highlights the emergence and significance of evolving intelligent systems in fuzzy modeling and control, emphasizing their advantages in handling nonstationary environments. Key challenges and future directions are discussed, including safety, interpretability, and principled structural evolution.

\end{abstract}


\section{Introduction}

Research in fuzzy modeling, control, and applications has grown rapidly since Zadeh's seminal work in 1965 \cite{zadeh1965}, evolving into a vast and multifaceted field. It includes key theoretical advances and a wide range of applications in engineering, industry, mechatronics, computer science, information systems, and beyond. This body of work has offered valuable opportunities to refine foundational concepts and increase the visibility of practical implementations. The established theoretical results, along with numerous successful applications—highlighting performance, smoothness, robustness, and interpretability in support of human decision-making—stand as evidence of the widespread acceptance and practical viability of fuzzy models and fuzzy control systems.

Despite evolving attitudes toward fuzzy systems, particularly within the academic community, debates have persisted between proponents and critics. As summarized by Belohlavek \cite{belohlavek2017}: (i) Lindley (1987) argued that anything achievable through fuzzy logic, belief functions, upper and lower probabilities, or other alternatives to probability could be done better using probability theory; (ii) Cheeseman (1986) similarly claimed that everything achievable with fuzzy logic is better addressed probabilistically; and (iii) in 1972, Kalman—one of the pioneers of modern system modeling and control—asserted that fuzzy logic was pragmatically unconvincing due to the lack of evidence it could solve important problems.

Interestingly, Kosko \cite{kosko1992} developed and compared a fuzzy control system with a Kalman filter-based control system for real-time target tracking. Simulation results suggested that, in many cases, fuzzy controllers may provide a robust and computationally efficient alternative to both linear and nonlinear (extended) Kalman filter approaches in real-time control—even when accurate input-output differential or difference equation models are available. Following this work—and even throughout the late 1980s, 1990s, and early 2000s—a growing number of studies confirmed similar findings, demonstrating that fuzzy control methods can offer competitive or superior performance, along with stability and robustness, in a variety of real-time, nonlinear, and uncertain environments.

In the preface of \cite{ying2000}, Zadeh wrote: ``What is not fully recognized, however, is that fuzzy control and conventional crisp control are, for the most part, complementary rather than competitive. Thus, fuzzy control is rule-based whereas conventional control is differential-equation-based; fuzzy control is task-oriented whereas conventional control is set-point-oriented; and conventional control is model-based whereas, in the case of fuzzy control, what suffices is a linguistic, rule-based description of the model. Today we see more clearly that fundamentally, conventional control is measurement-based whereas fuzzy control is perception-based. In this sense, the role model for fuzzy control is the remarkable human capability to perform a wide variety of tasks without any measurements and any computations. A canonical example of such tasks is that of driving a car in city traffic. Classical control provides no methods for automation of tasks of this type." This perspective highlights the foundational role of fuzzy sets in representing expert knowledge and supporting intelligent, flexible control strategies. 

In recent years, the integration of machine learning and knowledge discovery techniques into fuzzy models and controllers has further advanced the expert-based paradigm of the 1990s. This progress has led to the development of novel and adaptive approaches that merge rule-based reasoning with data-driven learning and model-based control design. In parallel, formal tools from optimization and dynamical systems theory—such as regularization, constrained loss functions, Lyapunov stability analysis, linear matrix inequalities, H-infinity control, model predictive control, sector nonlinearity, and sum-of-squares programming—have been increasingly adopted to provide theoretical guarantees \cite{Tanaka2004, Leite2015, Skrjanc2019}, further enriching fuzzy control with mathematical rigor and strengthening its alignment with contemporary control theories.

Current developments in fuzzy modeling and control have focused on strengthening their capacity to manage complexity, uncertainty, and imprecision in dynamic environments. Emphasis has shifted toward greater autonomy in model construction and adaptation, with algorithms capable of extracting and updating fuzzy rules, membership functions, and locally valid equations directly from data. These approaches significantly reduce, or even eliminate, the dependence on manual expert input, allowing more scalable and adaptive solutions aligned with current trends in modeling large datasets and data streams \cite{Skrjanc2019}. Furthermore, the integration of fuzzy systems with machine learning and data science has led to hybrid frameworks that support online learning, generalization, and better integration with real-world applications.

The next sections revisit the area of evolving fuzzy systems—an established branch of fuzzy modeling and control that focuses on systems capable of incrementally adapting their structure, functionality, and knowledge in response to nonstationary data streams \cite{Skrjanc2019, kasabov2002, plamen2002, plamen2004, Sayed2012}. The discussion begins with a brief overview of classical paradigms for modeling and controlling complex systems, followed by the notion of adaptive modeling and control. It then highlights the key ideas and contributions that evolving fuzzy systems have brought to the broader field over the past decade.

\section{Modeling and Control of Complex Systems}

The modeling and design of control systems for complex dynamical processes remain difficult and challenging tasks \cite{khalil2002, CoppHespanha2017}. While a rich and well-established body of theory exists for linear systems, the same level of maturity has not yet been achieved for nonlinear control. In most cases, analytic solutions to nonlinear control problems are unavailable, requiring approximate or heuristic approaches.

Conventional design methods for complex nonlinear systems are typically model-based and often laborious, involving subjective steps informed by prior experience or simulation-based tuning. A common practice is to approximate a nonlinear control law using multiple linear controllers, each designed for a specific region of the operating space. In each region, the system dynamics is assumed to be approximately linear. This partition-based (or granular) approach becomes necessary when a single linear model fails to adequately represent the system's global behavior. The nonlinear control strategy is then synthesized by switching or interpolating among the local controllers based on the current state of the system. However, this approach is inherently sensitive to modeling inaccuracies. Any significant mismatch between the model and the actual plant, or the presence of unmodeled and time-varying (evolving) dynamics, can severely degrade performance and even lead to instability. Importantly, such limitations may persist despite rigorous offline controller design.

To ensure robust performance, controllers are generally expected to tolerate a certain degree of modeling uncertainty and external disturbances \cite{bernal2022}. However, robustness is often achieved at the expense of optimal closed-loop performance. A promising alternative lies in control systems capable of learning from experience in real time. By continuously adapting to previously unmodeled and time-varying (evolving) dynamics, these systems offer the potential to reconcile robustness with high performance—ultimately motivating the development of evolving control strategies.

\subsection{Classical Adaptive Control and Its Limitations}

Adaptive modeling and control systems adjust their parameters in response to changes in the dynamical behavior of the controlled process. The concept originated in the 1950s \cite{aseltine1958} and was soon formalized within the framework of stochastic systems by Feldbaum \cite{feldbaum1960}, whose work is widely regarded as the first theoretical foundation of adaptive control. Additional frameworks emerged in the early 1960s, notably the work of Mishkin and Braun \cite{mishkin1961}, who provided a systematic treatment of adaptive control principles and motivated their application in engineering contexts. Around the same period, further developments and perspectives began to appear in the literature, such as the contribution by Truxal \cite{truxal1963}, reflecting the growing interest in adaptive techniques.

Adaptive modeling and control methods monitor the input/output behavior of a plant to identify, either explicitly or implicitly, the parameters of the design model and adjust the controller parameters accordingly to meet the specified performance. An adaptive system attempts to revise these parameters whenever the plant's behavior changes significantly. If the dynamical characteristics of the process vary over its operational range, the control system may be required to adapt continuously. In essence, adaptive modeling and control combine an online parameter estimator with a control law to regulate or compensate classes of plants whose parameters are either initially unknown or vary over time in unpredictable ways. The particular choices of estimator, control law, and their integration define different classes of adaptive control schemes \cite{ioannou2006}.

\begin{figure}[h]
\centering
\includegraphics[width=1\linewidth]{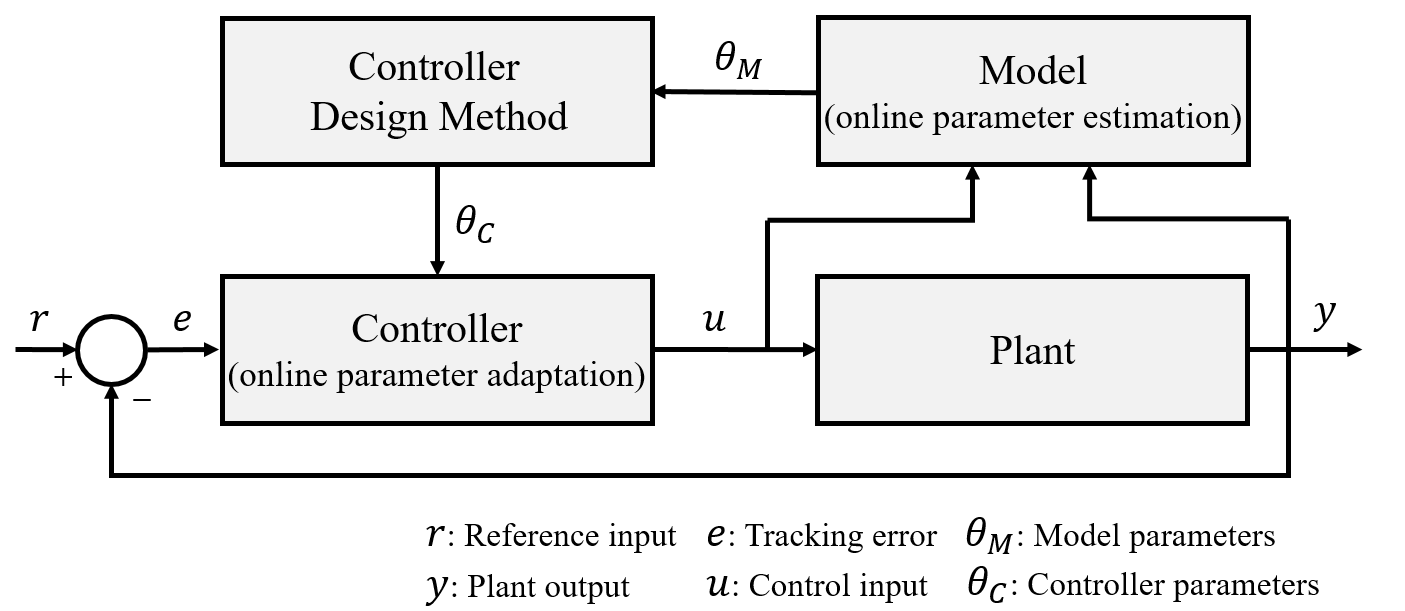}
\caption{Indirect adaptive control scheme}
\label{fig1}
\end{figure}

The main approaches to adaptive control differ in whether the controller parameters are adapted directly or indirectly. In direct adaptive control, the controller parameters are adjusted directly on the basis the observed system behavior. In indirect adaptive control, the plant model parameters are first estimated online, and the controller is subsequently updated as a function of these estimates (Fig. \ref{fig1}). 

Direct adaptive control was first introduced in the context of model reference adaptive systems for aircraft control \cite{whitaker1958}, while indirect adaptive control emerged in digital process control applications \cite{kalman1958}. These adaptation strategies underpin classical schemes such as Model Reference Adaptive Control (MRAC) and Self-Tuning Regulators (STR) \cite{Skrjanc2003, Skrjanc2003b}, which remain central to online adaptive control. MRAC defines the desired closed-loop behavior via an explicit reference model and adapts the controller to minimize the model-following error, whereas STR estimates plant parameters online and recomputes the controller using techniques such as pole placement or optimal control synthesis.

The use of adaptive control is based on the assumption that, for any possible values of the coefficients of the plant model, there exists a controller—with fixed structure and complexity—capable of meeting the design specifications through appropriate tuning of its parameters \cite{landau2011}. In this context, the task of adaptation is to determine suitable values for the controller parameters. It is worth noting that conventional controller design relies on an offline mathematical model of the process. Once an adequate model is obtained, established design methods are used to synthesize a controller that meets the required performance specifications. In contrast to this static procedure, adaptive control systems aim to adjust model and controller parameters online, while keeping their structures fixed by design. These adjustments are driven by real-time input-output data from the process. 

Ultimately, in conventional design, both the model and controller structures—as well as their coefficients and gains—are fixed during operation. Adaptive systems allow real-time adjustment of model coefficients and controller gains but still assume fixed structures. The simultaneous adaptation of both the structure and parameters of the model and controller remains an open challenge. As highlighted by Annaswamy and Fradkov \cite{annaswamy2022}, adaptive control remains primarily concerned with parameter adaptation—that is, the tuning of model coefficients and controller gains. This limitation has motivated the development of evolving control schemes capable of structural and parametric learning from data streams.

\subsection{Toward Evolving Modeling and Fuzzy Control}

\begin{figure}[h]
\centering
\includegraphics[width=0.98\linewidth]{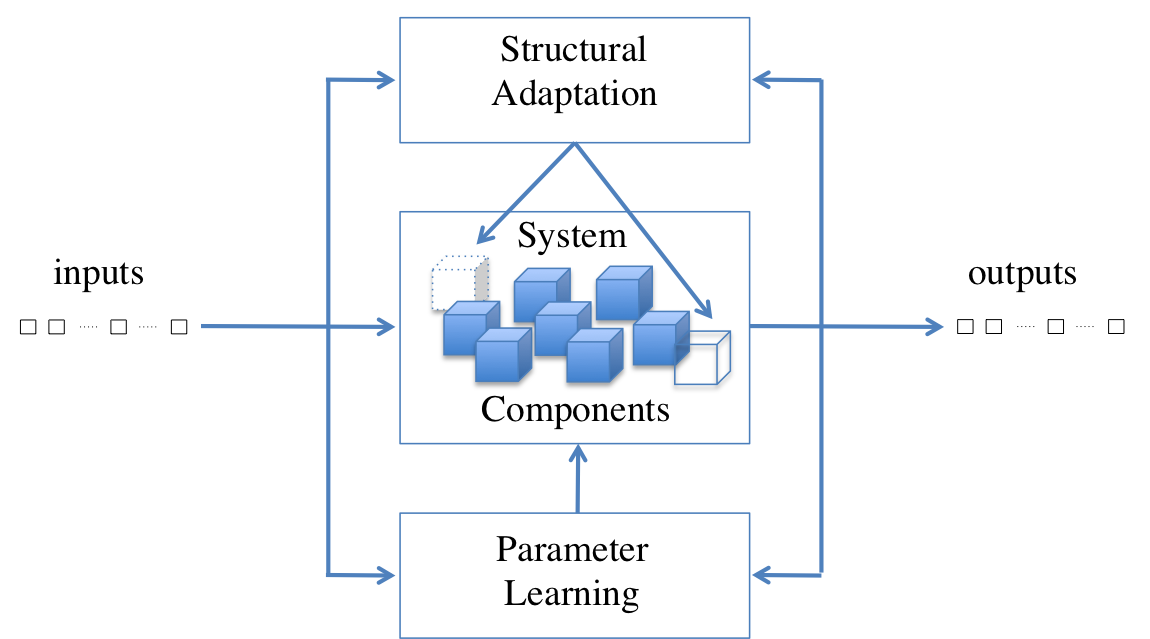}
\caption{Evolving system—plant model or controller: Online adaptation of both structure and parameters}
\label{fig2}
\end{figure}

Evolving fuzzy systems represent a major shift in the field of adaptation, learning, and self-organizing systems, with impact extending well beyond fuzzy modeling and control, including online classification, clustering, forecasting, and decision-making in dynamic environments \cite{Skrjanc2019}. In contrast to conventional modeling and control methods—which require an offline design or training phase—evolving fuzzy systems simultaneously adapt both their structure and parameters (Fig. \ref{fig2}). They emphasize incremental learning and self-development, can operate entirely online, start from scratch without prior process knowledge, and allow human knowledge to be incorporated at any stage. The general scheme illustrated in Fig. \ref{fig2} is applicable to both evolving plant models and controllers, provided that the appropriate input-output signals are considered. A practical example employing both an evolving Takagi–Sugeno \cite{takagi1985} fuzzy model of the process and an evolving fuzzy controller, within a parallel distributed compensation (PDC) \cite{Tanaka2004} strategy, was first presented in \cite{Leite2015}, where bounded control inputs and Lyapunov stability were formally ensured.

Quoting the pioneers of the area \cite{plamen2008}: ``The newly emerging concept of dynamically evolving structures, which was, to some extent, already applied to neural networks \cite{fritzke1994}, brought a powerful new concept of evolving fuzzy systems (EFS). EFS combine: (i) the interpolation abilities of fuzzy systems; (ii) their flexibility; (iii) the linguistic interpretability of fuzzy systems; and (iv) the adaptive feature of online learning techniques. This new topic was introduced during the last decade \cite{kasabov2002, plamen2002, plamen2004}, and quickly numerous applications to problems of modeling, control, prediction, classification, and data processing in a dynamically changing and evolving environment were also developed, including some successful industrial applications \cite{kordon2006}."

The terms adaptation, learning system, and self-organization often appear in the literature as synonymous. Early work views an adaptive system as one that is insensitive to changes in its environment, that is, one that performs acceptably well over a range of inputs \cite{zadeh1963}. In contrast, a learning system is one that operates satisfactorily under changing environmental conditions in which an adaptive system does not improve its performance \cite{hill1965}. Self-organization refers to a process in which the structure and behavior of a system emerge from interactions among its components. This contrasts with conventional modeling and control systems, which rely on centralized or fixed hierarchical structures. Self-organization enables more flexible, adaptable, and resilient systems capable of shaping their own behavior \cite{saridis1975}.

Current data-driven learning models and control methods typically require an offline training phase, often based on input-output datasets generated by applying a wide range of input conditions in closed loop. This process is frequently infeasible, especially for originally unstable systems. Moreover, these methods lack the capability to operate from scratch and to perform concurrent, online adjustment of both system structure and parameters—a key requirement for fully self-organizing and evolving systems.

\subsection{Open Challenges and Paths Forward}

Closing the loop with a controller that evolves from scratch is risky, especially in open-loop unstable systems or those involving safety concerns. While evolving fuzzy systems offer the potential to self-develop from minimal prior knowledge, their effective use in practice requires guarantees of safe operation—especially during the early stages of learning. To this end, possible strategies include: (i) using initial conservative controllers (fallback control) while the evolving model and controller learn cooperatively through observation; (ii) incorporating human-in-the-loop supervision; (iii) restricting exploration to known safe regions of the input space; and (iv) implementing bounded output policies that ensure safe actuator behavior. Evolving fuzzy control should balance autonomy with safety-aware initialization and updates to maintain system stability and operational integrity.

Beyond safety, interpretability remains a key requirement for the practical adoption of evolving fuzzy controllers. Future research should address how to evolve models and controllers in a controlled manner. Structural changes—such as the addition of new fuzzy rules—should be validated in shadow or standby mode via separate computational simulations, or be subject to human review before activation. Risk metrics and domain-specific safety constraints should guide the evolution process to protect critical state variables. Additionally, robustness to uncertainty and resilience to anomalies are essential to prevent overreaction and maintain system stability.

\section{Conclusion}

This paper revisited the fundamental concepts of adaptive modeling and control, highlighting their capabilities and limitations in the face of nonstationary and complex environments. While conventional adaptive systems support online parameter adjustment, they operate under fixed model and controller structures, which limits their ability to handle complex, evolving processes. Evolving fuzzy systems fill this gap by supporting both parametric and structural adaptation from data streams, often without requiring prior process knowledge. Their capacity for incremental learning and integration of expert opinion makes them a promising path forward. However, applying these systems in real-world scenarios—particularly those inherently unstable or involving safety risks—demands additional mechanisms to ensure safe learning, interpretability, and resilience. As such, future efforts should balance structural adaptability and domain-aware validation.

\section*{Acknowledgment}

The first author acknowledges the Ministry of Culture and Science of the State of North Rhyne-Westphalia for grant no. NW21-059D, SAIL project. The last author is grateful to the Brazilian National Council for Scientific and Technological Development for grant 302467/2019-0.

\vspace{1cm}

\color{black}


\begin{thebibliography}{00}

\bibitem{zadeh1965} L. A. Zadeh. Fuzzy sets. Information and Control, 8(3), 338–353, 1965.

\bibitem{belohlavek2017} R. Belohlavek, J. Dauben, J. Klir, Fuzzy Logic and Mathematics: A Historical Perspective. New York, NY: Oxford, 2017.

\bibitem{kosko1992} B. Kosko, Fuzzy Logic: A Dynamical System Approach to Machine Intelligence, Englewood Cliffs, NJ: Prentice-Hall, 1992.

\bibitem{ying2000} H. Ying, Fuzzy Control and Modeling: Analytical Foundations and Applications. New York, NY: IEEE Press, 2000.

\bibitem{Tanaka2004} K. Tanaka, H. O. Wang, Fuzzy Control Systems Design and Analysis: A Linear Matrix Inequality Approach. J. Wiley \& Sons, Hoboken, 2004.

\bibitem{Leite2015} D. Leite, R. Palhares, V. Campos, F. Gomide, Evolving granular fuzzy model-based control of nonlinear dynamic systems. IEEE Transactions on Fuzzy Systems, 23(4), 923-938, 2015.

\bibitem{Skrjanc2019} I. Škrjanc, J. A. Iglesias, A. Sanchis, D. Leite, E. Lughofer, F. Gomide, Evolving fuzzy and neuro-fuzzy approaches in clustering, regression, identification, and classification: A survey. Information Sciences, 490, 344-368, 2019.

\bibitem{kasabov2002} N. Kasabov, Q. Song, DENFIS: Dynamic evolving neural-fuzzy inference system and its application for time-series prediction. IEEE Transactions on Fuzzy Systems, 10, 144–154, 2002.

\bibitem{plamen2002} P. Angelov, Evolving Rule-based Models: A Tool for Design of Flexible Adaptive Systems. Heidelberg: Springer, 2002.

\bibitem{plamen2004} P. Angelov, D. Filev, An approach to online identification of Takagi–Sugeno fuzzy models. IEEE Transactions on Systems, Man, and Cybernetics B, 34, 484–498, 2004.

\bibitem{Sayed2012} M. Sayed-Mouchaweh, E. Lughofer (Eds.), Learning in Non-Stationary Environments: Methods and Applications. Springer, New York, 2012.

\bibitem{khalil2002} H. Khalil, Nonlinear Systems. Upper Saddle River, NJ: Prentice Hall, 2002.

\bibitem{CoppHespanha2017} D. A. Copp, J. P. Hespanha, Simultaneous nonlinear model predictive control and state estimation. Automatica, 77, 143–154, 2017.

\bibitem{bernal2022} M. Bernal, A. Sala, Z. Lendek, T. Guerra, Analysis and Synthesis of Nonlinear Control Systems. Cham, CH: Springer, 2022.

\bibitem{aseltine1958} A. Aseltine, R. Mancini, C. Saturne, A survey of adaptive control systems. IRE Transactions on Automatic Control, 6, 102–108, 1958.

\bibitem{feldbaum1960} A. Feldbaum, Dual control theory. Automation and Remote Control, 21, 1240–1249, 1960.

\bibitem{mishkin1961} E. Mishkin, L. Brown, Adaptive Control Systems. New York, NY: McGraw-Hill, 1961.

\bibitem{truxal1963} J. Truxal, Adaptive control. Proceedings of IFAC, 1, 386–392, 1963.

\bibitem{ioannou2006} P. Ioannou, B. Fidan, Adaptive Control Tutorial. Philadelphia, PA: SIAM, 2006.

\bibitem{whitaker1958} H. Whitaker, J. Yamron, A. Kezer, Design of a model-reference adaptive control system for aircraft. Tech. Report R-164, MIT Instrumentation Laboratory, Cambridge, MA, 1958.

\bibitem{kalman1958} R. Kalman, Design of self-optimizing control systems. Transactions of ASME Journal of Basic Engineering, 80, 468–478, 1958.

\bibitem{Skrjanc2003} S. Blažič, I. Škrjanc, D. Matko, Globally stable direct fuzzy model reference adaptive control. Fuzzy Sets and Systems, 139(1), 3–33, 2003.

\bibitem{Skrjanc2003b} I. Škrjanc, S. Blažič, D. Matko, Model-reference fuzzy adaptive control as a framework for nonlinear system control. Journal of Intelligent \& Robotic Systems, 36(3), 331–347, 2003.

\bibitem{landau2011} I. Landau, R. Lozano, M. Saad, A. Karimi, Adaptive Control: Algorithms, Analysis and Applications. New York, NY: Springer, 2011.

\bibitem{annaswamy2022} A. Annaswamy, A. Fradkov, A historical perspective of adaptive control and learning. https://doi.org/10.48550/arXiv.2108.11336, 2022.

\bibitem{takagi1985} T. Takagi, M. Sugeno, Fuzzy identification of systems and its applications to modeling and control. IEEE Transactions on Systems, Man, and Cybernetics, 15(1), 116–132, 1985.

\bibitem{plamen2008} P. Angelov, N. Kasabov, D. Filev, Guest editorial evolving fuzzy systems: Preface to the special section. IEEE Transactions on Fuzzy Systems, 16, 1391–1392, 2008.

\bibitem{fritzke1994} B. Fritzke, Growing cell structures—A self-organizing network for unsupervised and supervised learning. Neural Netw, 7, 1441–1460, 1994.

\bibitem{kordon2006} A. Kordon, Inferential sensors as potential application area of intelligent evolving systems. Proceedings of IEEE International Symposium on Evolving Fuzzy Systems, UK, 2006.

\bibitem{zadeh1963} L. Zadeh, On the definition of adaptivity. Proceedings of the IEEE, 51, 469–470, 1963.

\bibitem{hill1965} J. Hill, G. McMurthy, K. Fu, A computer-simulated on-line experiment in learning control systems. Simulation, 4, 104–116, 1965.

\bibitem{saridis1975} G. Saridis, Self-organizing control and applications to trainable manipulators and learning prostheses. IFAC Proceedings, 8, 632–640, 1975.





\end{thebibliography}
\end{document}